# Modulation and Switching Architecture Performances for Frequency Up-conversion of Complex-Modulated Data Signals based on a SOA-MZI Photonic Sampling Mixer

Dimitrios Kastritsis, Thierry Rampone, Kyriakos E. Zoiros, Ammar Sharaiha

***Abstract*—A theoretical and experimental performance analysis of a Semiconductor Optical Amplifier – Mach-Zehnder Interferometer (SOA-MZI) photonic sampling mixer used as a frequency up-converter is presented employing Switching and Modulation architectures. An active mode-locked laser, generating 2 ps-width pulses at a repetition rate equal to 10 GHz, is used as a sampling source. An optical carrier intensity modulated by a sinusoidal signal at 1 GHz is up-converted to 9 GHz and 39 GHz. High Conversion Gains (CGs) of about 15 dB are demonstrated for the frequency conversion to 9 GHz using both architectures, whereas up to 4 dB and 9 dB for the conversion to 39 GHz employing Switching and Modulation architectures, respectively. Small-signal equations for the up-converted signal in both architectures are formulated and developed, which permit to quantify the CG from closed-form expressions. The numerically calculated CG values are in very good agreement with those obtained experimentally. The validated equations are subsequently employed to explain the performance differences between the two architectures in terms of the CG. Furthermore, signals modulated by QPSK and 16-QAM complex modulation formats at different baud rates are up-converted from 750 MHz to 9.25 GHz and 39.75 GHz and their Error Vector Magnitude is evaluated and compared. The maximum bit rate that meets the Forward Error Correction (FEC) limit is achieved using the Modulation architecture. It is 1 Gbps and 512 Mbps for QPSK and 16-QAM modulations, respectively.**

## I. Introduction

Radio over Fiber (RoF) systems have attracted significant attention by enabling the transmission of microwave frequency signals via optical fiber, thus benefiting from the advantages of low loss transmission, low weight, wide bandwidth and immunity to electromagnetic interference compared to an all-electrical transmission. The potential range of RoF applications includes, but is not limited to, cellular communications, satellite communications, Wireless Fidelity (WiFi), phased array and photonics-assisted radar systems [1]. In this framework, and in order to expand the potentiality and flexibility of such a system, the development of a subsystem in the optical layer for optical generation of millimeter-wave (mm-wave) signals through optical mixing is crucial [1]-[5].

There are three fundamental characteristics of an optical mixer: the nonlinear medium that is exploited for the implementation of the optical mixer, the physical phenomenon used for the frequency conversion and the device that performs the function of mixing. The fused silica in a Highly Nonlinear Dispersion Shifted Fibers (HNL-DSF) can be used for optical mixing exploiting either the Cross-Phase Modulation (XPM) or the Four-Wave Mixing (FWM) phenomena [1]. The main restriction of HNL is the need of large fiber length and high power. High performance Mach-Zehnder modulators (MZM) with Lithium-Niobate material (LiNbO3) that exploit Pockels electro-optic effect have been extensively used for the purpose of frequency conversion [1]. Another possibility is the Electro-Absorption Modulator utilizing Cross-Absorption Modulation (XAM) phenomenon in a semiconductor material [6][7]. Finally, photodiodes and lately UniTravelling Carrier PhotoDiodes (UTC-PD) have also been employed by exploiting the nonlinear current response when the incident power is high enough and the reverse bias of photodiode is low enough [8]. Semiconductor Optical Amplifiers (SOA) exploiting the nonlinear phenomena of XGM (Cross-Gain Modulation) [9] and XPM [10]-[17] can be equally used for mixing in the form of a stand-alone device or embedded in an interferometric arrangement such as a Semiconductor Optical Amplifier-Mach-Zehnder Interferometer (SOA-MZI). Lately, SOA-MZI-based frequency mixer was successfully used to generate a 100 GHz signal through up-conversion of 1 GHz signal [11]. SOA and SOA-MZI exhibit all the advantages mentioned in the previous paragraph while being an all-optical solution. In addition, because they both are active modules, they also provide a

This work is part of a Ph. D. Thesis under co-supervising scheme ‘Cotutelle’ between ENIB in France and DUTH in Greece with the support of Brest Métropole in the frame of ARED ‘Choraal’ project. It is also supported by the French state, Brittany region and FEDER in the frame of CPER SOPHIE-Photonique-ATOM.

D. Kastritsis is both with Lab-STICC UMR CNRS 6285, École Nationale d’Ingénieurs de Brest, 29238 Brest, France, and with the Department of Electrical and Computer Engineering, Lightwave Communications Research Group, Democritus University of Thrace, 67100 Xanthi, Greece (e-mail: dimitrios.kastritsis@enib.fr).

T. Rampone and A. Sharaiha are with Lab-STICC UMR CNRS 6285, École Nationale d’Ingénieurs de Brest, 29238 Brest, France (e-mail: thierry.rampone@enib.fr; ammar.sharaiha@enib.fr).

K. E. Zoiros is with the Department of Electrical and Computer Engineering, Lightwave Communications Research Group, Democritus University of Thrace, 67100 Xanthi, Greece (e-mail: kzoiros@ee.duth.gr).

Conversion Gain (*CG*) as opposed to the conversion loss suffered by the previously reported passive photonic mixers [17]. The main disadvantage of XGM-based mixing techniques is the limited extinction ratio of the converted signal. By comparison, a SOA-MZI provides large extinction ratio while lowering the required optical input power due to its interferometric structure [12]-[13]. A comprehensive comparative analysis of alternative up-conversion configurations employed in RoF systems is made in [1].

All-optical mixing using a SOA-MZI has been experimentally demonstrated in [10], where frequency up-conversion from 2.5 GHz to 32.5 GHz was demonstrated with a high conversion gain of 6 dB. A SOA-MZI combined with a mode-locked laser's pulse train, instead of a two-tone signal produced either by an electro-optic modulator, as in [10], or an MZM, as in [12], provides increased flexibility as it can be used to up- or down-convert a signal at multiple target frequencies simultaneously [13]-[14].

Recently, a similar SOA-MZI-based system [14] reported *CG* for multiple target frequencies. Specifically, a signal at 0.5 GHz was up-converted simultaneously at frequencies between 8.3 GHz, which was adjacent to the 1st harmonic of the mode-locked laser, and 30.5 GHz, which was adjacent to the 4th harmonic of the mode-locked laser. The conversion gain achieved in this range of frequencies was reduced from 15.5 dB to -9.5 dB.

In our recent works [15]-[16], we have experimentally demonstrated a new Modulation architecture and we have compared its performance in terms of the *CG* against the established Switching architecture, both based on the SOA-MZI photonic sampler. In [15], a very high *CG* up to 22 dB was achieved, where a 1 GHz was up-converted to 9 GHz, which was adjacent to the 1st harmonic of the mode-locked laser.

In this paper, we extend the comparison by presenting a small-signal analysis for the SOA-MZI photonic sampling mixer. In this context, we derive an analytical expression for the *CG* and provide a qualitative insight into the trends of *CG* results for the two architectures. The performance in terms of *CG* is evaluated and the analytical expression is validated by comparison to measured *CG* values. The highest, to our knowledge, *CG*, is achieved with the Modulation architecture of the SOA-MZI photonic mixer for up-conversion of a signal from 1 GHz to 39 GHz, which is adjacent to the 4th harmonic of the mode locked laser. Moreover, we evaluate the performance of up-conversion using QPSK and 16-QAM complex modulated signals which is done, to the best of our knowledge, for the first time for the Modulation architecture. The remainder of the paper is organized as follows. In Section II, we describe the principle of operation for Switching and Modulation architectures. In Section III, we provide a small-signal analysis of the SOA-MZI sampling mixer. In Section IV, we define the conditions of the conducted experiments and give static and CG experimental results. In Section V, we present results for the frequency up-conversion of complex-modulated data by comparing the two architectures against the Error Vector Magnitude (EVM). Finally, Section VI contains the conclusions reached from this work.

## II. Principle of Operation of SOA-MZI Sampling Mixer

The SOA-MZI is an optically controlled device, in which a SOA is placed at each arm of a MZI. In the generic case there are two inputs, $P_{ctrl}$ at $\lambda_{ctrl}$ and $P_{in}$ at $\lambda_{in}$, and two output signals, $P_I$ and $P_J$, where '$P$' denotes power. $P_{ctrl}$ governs through XPM the SOA-MZI behavior by defining the fraction of the amplified input signal that exits from the upper and lower output ports. Fig. 1 demonstrates the SOA-MZI's photonic sampling mixer principle of operation for the Switching and the Modulation architectures. We inject two signals at the SOA-MZI input: the sampling pulse train signal, which consists of ultra-short continuous (clock) pulses with repetition frequency $f_{ck}$, and the signal to be sampled, which consists of a sinusoidal signal at frequency $f_{dat}$ superimposed on a Continuous Wave (CW) signal. Fig.1 (a) illustrates the Switching architecture, in which the signal to be sampled is switched on and off by the sampling pulse train as follows. The signal to be sampled is divided at the SOA-MZI middle input port into two identical copies by a 3 dB optical coupler. At time instants at which no sampling pulse is present, the signal to be sampled is amplified by the two SOAs and exits from the upper output port. At time instants at which a sampling pulse is present, a phase shift is induced in the upper copy of the input signal due to XPM phenomenon and a portion of the signal to be sampled appears amplified at lower output port, thus resulting in the sampling of the sinusoidal signal. Provided that the peak power of the pulses causes a differential phase shift of π, the entire signal to be sampled emerges amplified at the lower output port. The two phase shifters $\Phi_{01}$ and $\Phi_{02}$ introduce a static phase shift between the two arms in order to compensate for asymmetries in the couplers or between the two SOAs. The idea for realizing a sampling mixer is to exploit the frequency response of the sampling process to achieve frequency conversion [13]-[19]. As we can see from the power spectrum of the sampled signal obtained after optical filtering centered at $\lambda_{in}$, the frequency content of the signal to be sampled is replicated from $f_{dat}$ to $f_{cki} \pm f_{dat}$, where $f_{cki} \equiv i f_{ck}$, $i \in \mathbb{N}^+$. Fig.1(b) demonstrates the principle of operation for the Modulation architecture in which the sampling pulse train is modulated by the sinusoidal signal to be sampled. More specifically, it is now the sampling pulsed signal that is divided into two identical copies by the 3 dB optical coupler at the middle input port. In this case, the signal to be sampled causes a phase shift in the upper copy of the sampling pulse train. This results in a continuous variation of the portion of the amplified sampling signal that appears at the upper and lower output, i.e. in a modulation of the sampling pulse train by the signal to be sampled. Again the frequency content of the signal power to be sampled after the optical filtering at $\lambda_{in}$ is replicated from $f_{dat}$ to $i f_{ck} \pm f_{dat}$.

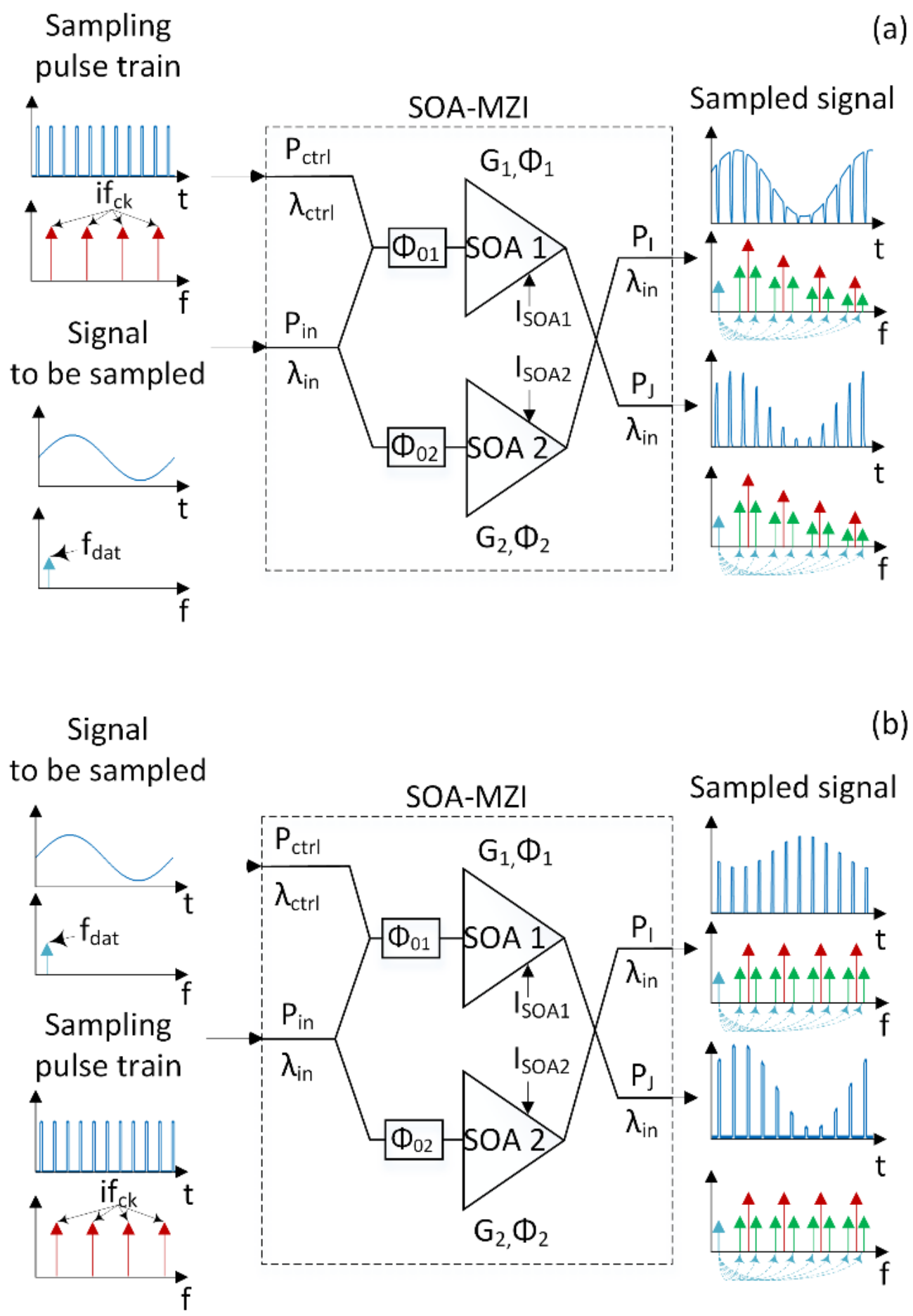

Fig. 1. SOA-MZI (a) Switching and (b) Modulation architectures.

## III. Small-Signal Analysis of SOA-MZI Sampling Mixer

A small-signal analysis of the SOA-MZI sampling mixer is formulated by expressing each time-dependent parameter as the sum of a steady state term $\bar{X}$ and a first-order perturbation term $\delta x$ at one or more angular frequencies [20]-[23]:

$$X(t) = \bar{X} + \delta x \tag{1}$$

In our case one input is a sampling pulse train that is modeled as an infinite sum of harmonics.

### A. Modeling in presence of a sampling pulse train

Using the notation (1) and the index $a \in \{s, m\}$ to discriminate between the (*s*)witching and (*m*)odulation architectures, the power of the sampling signal can be written as the sum of the average power $\bar{P}_{ck,a}$ and the perturbation term $\delta p_{cki,a}$, which represents the sinusoidal variation of the $i^{th}$ harmonic at angular frequency $\omega_{cki}$:

$$P_{ck,a} = \bar{P}_{ck,a} + \delta p_{cki,a}$$

$$= \bar{P}_{ck,a} + \frac{1}{2}\sum_{i\in\mathbb{N}^+}\left(p_{cki,a}e^{j\omega_{cki}t} + p^*_{cki,a}e^{-j\omega_{cki}t}\right) \tag{2}$$

Likewise, the input power of the signal to be sampled, $P_{dat,a}$, is the sum of an average power $\bar{P}_{dat,a}$ and a sinusoidal variation $\delta p_{dat,a}$ at the angular frequency $\omega_{dat}$:

$$P_{dat,a} = \bar{P}_{dat,a} + \delta p_{dat,a}$$

$$= \bar{P}_{dat,a} + \frac{1}{2}\left(p_{dat,a}e^{j\omega_{dat}t} + p^*_{dat,a}e^{-j\omega_{dat}t}\right) \tag{3}$$

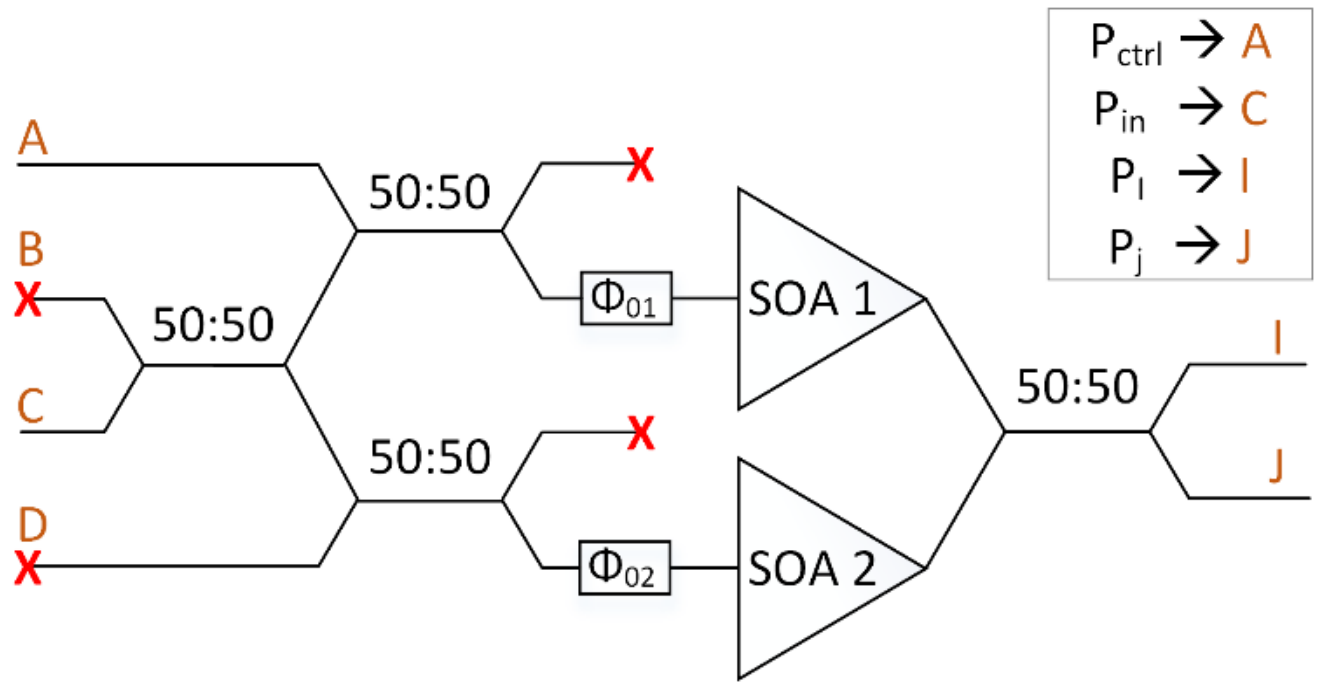

Fig. 2. SOA-MZI layout of CIP 40G-2R2-ORP device.

Fig.2 depicts the layout of the SOA-MZI (CIP, model 40G-2R2-ORP) used in the experiments. With reference to the input and output powers which have been defined in the previous section, $P_{in}$ and $P_{ctrl}$ enter the interferometric module from port C and A, while $P_I$ and $P_J$ emerge from ports I and J, respectively. Ports B and D have not been used in this work.

At SOA-MZI input port C, the incoming power is decomposed as:

$$P_{in} = \bar{P}_{in} + \delta p_{in} \tag{4}$$

Similarly, introducing the index $k \in \{1, 2\}$ to discriminate between SOA1 and SOA2, we get the following equation for the carrier density, $N_k$ (the SOA is modelled as a one section device):

$$N_k = \bar{N}_k + \delta n_k$$

$$= \bar{N}_k + \frac{1}{2}\sum_F\left(n_{k,\omega_F}e^{j\omega_F t} + n^*_{k,\omega_F}e^{-j\omega_F t}\right) \tag{5}$$

where $\omega_F$ represents the linear combination of the angular frequency components of all optical signals traversing SOAk. In case of SOA1, these include the sampling signal and the signal to be sampled. In case of SOA2, only one of the two optical signals passes through, i.e. either the signal to be sampled, thus $\omega_F = \omega_{dat}$ for the Switching architecture, or the sampling signal, thus $\omega_F = \omega_{cki}$ for the Modulation architecture.

The SOAs time-dependent gain and phase properties can be developed as follows:

$$G_k = \bar{G}_k + \delta g_k = \bar{G}_k + \delta n_k \frac{\partial G_k}{\partial N} \tag{6}$$

$$\Phi_k = \bar{\Phi}_k + \delta \varphi_k = \bar{\Phi}_k + \delta n_k \frac{\partial \Phi_k}{\partial N} \tag{7}$$

At SOA-MZI output port J the optical power is given by [23]:

$$P_J = \frac{1}{8} P_{in}\left(G_1 + G_2 - 2\sqrt{G_1 G_2}\cos(\Phi_1 - \Phi_2 + \Phi_0)\right) \tag{8}$$

where $\Phi_0 = \Phi_{01} - \Phi_{02}$. The static phase shift $\Phi_0$ introduced by the optical phase shifters will be considered equal to 0 in our operating conditions. The factor 1/4 in [23] is modified to 1/8 because of the additional 3 dB coupler used to inject the signal at the upper and the lower SOA-MZI arms.

Now applying to (8) a first-order Taylor series approximation around $[\delta g_1, \delta g_2, \delta\varphi_1, \delta\varphi_2]$ = [0, 0, 0, 0] and introducing, (6) and (7), we get (9). Developing (9), we get the average output power $\bar{P}_J$ and variation power $\delta p_J$:

$$\bar{P}_J = \frac{1}{8} \bar{P}_{in}\left(\bar{G}_1 + \bar{G}_2 - 2\sqrt{\bar{G}_1 \bar{G}_2}\cos(\bar{\Phi}_1 - \bar{\Phi}_2)\right) \tag{10}$$

$$\delta p_J = \delta p_{in} \frac{\bar{P}_J}{\bar{P}_{in}} + \bar{P}_{in} \delta g\varphi + \delta p_{in} \delta g \varphi \tag{11}$$

where $\delta g\varphi$ is given by:

$$\delta g\varphi = \frac{1}{8}\left( \begin{array}{c} \delta g_1 + \delta g_2 \\ -\sqrt{\bar{G}_1 \bar{G}_2}\left(\frac{\delta g_1}{\bar{G}_1} + \frac{\delta g_2}{\bar{G}_2}\right)\cos(\bar{\Phi}_1 - \bar{\Phi}_2) \\ +2\sqrt{\bar{G}_1 \bar{G}_2}(\delta\varphi_1 - \delta\varphi_2)\sin(\bar{\Phi}_1 - \bar{\Phi}_2) \end{array} \right) \tag{12}$$

## *B. Up-converted small-signal terms*

We are going to develop (11) for the Modulation and Switching architectures in the following way. We are interested only in the combination of terms in (11) that are generating intermodulation products at frequencies $f_{cki} \pm f_{dat}$, while eliminating all other terms. For this purpose, firstly we eliminate the first term of (11) since we notice that it contains no intermodulation products but only the frequencies of the input signal at port C.

Secondly, we eliminate $\delta g_2$ and $\delta\varphi_2$ perturbation terms induced by SOA2 as we observe that they do not participate in the generation of $f_{cki} \pm f_{dat}$ intermodulation frequencies.

More analytically, for the Switching architecture, the only signal traversing SOA2 is $P_{dat}$, therefore $\delta g_2$ and $\delta\varphi_2$ variations are induced at frequency $f_{dat}$. As a result, for the second term of (11), the product of $\delta g_2$ and $\bar{P}_{in}$ as well as the product of $\delta\varphi_2$ and $\bar{P}_{in}$ correspond to variations at frequency $f_{dat}$. For the third term of (11), the product of $\delta g_2$ and $\delta p_{in}$ as well as the product of $\delta\varphi_2$ and $\delta p_{in}$ correspond to variations at frequency $2f_{dat}$.

Similarly, for the Modulation architecture, the only signal that passes through SOA2 is $P_{ck}$, therefore $\delta g_2$ and $\delta\varphi_2$ variations are induced at frequencies $f_{cki}$. As a result, for the second term of (11), the product of $\delta g_2$ and $\bar{P}_{in}$ as well as the product of $\delta\varphi_2$ and $\bar{P}_{in}$ correspond to variations at frequencies $f_{cki}$. For the third term of (11), the product of $\delta g_2$ and $\delta p_{in}$ as well as the product of $\delta\varphi_2$ and $\delta p_{in}$ correspond to variations at frequencies $2f_{cki}$.

Therefore, based on (11) and the previous analysis we can write for the intermodulation products at frequencies $f_{cki} \pm f_{dat}$:

$$\delta p_{J,cki\pm dat,a} = \frac{1}{8}(\bar{P}_{in}\,\delta g\varphi_1 + \delta p_{in}\delta g\varphi_1) \tag{13}$$

where $\delta g\varphi_1$ is given by:

$$\begin{aligned} \delta g\varphi_1 = \delta g_1 - \frac{\sqrt{\bar{G}_1 \bar{G}_2}}{\bar{G}_1}\cos(\bar{\Phi}_1 - \bar{\Phi}_2)\delta g_1 \\ + 2\sqrt{\bar{G}_1 \bar{G}_2}\sin(\bar{\Phi}_1 - \bar{\Phi}_2)\,\delta\varphi_1 \end{aligned} \tag{14}$$

Substituting $\delta g_1$, $\delta\varphi_1$ from (6) and (7) and $\delta n_k$ from (5), while taking into account that $P_{in} = P_{dat}$ for Switching and $P_{in} = P_{ck}$ for Modulation architectures, the optical powers at the up-conversion frequency $f_{cki-dat}$ are given by:

$$\begin{aligned} p_{J,cki-dat,s} = \frac{C_{OP_s}}{16}[p^*_{dat,s} n_{1,\omega_{cki},s} \\ + \bar{P}_{dat,s} 2n_{1,\omega_{cki-dat},s}] \end{aligned} \tag{15}$$

$$\begin{aligned} p_{J,cki-dat,m} = \frac{C_{OP_m}}{16}[p_{cki,m} n^*_{1,\omega_{dat},m} \\ + \bar{P}_{ck,m} 2n_{1,\omega_{cki-dat},m}] \end{aligned} \tag{16}$$

$$\begin{aligned} \bar{P}_J + \delta p_J = \frac{1}{8}(\bar{P}_{in} + \delta p_{in})\Bigg(\bar{G}_1 + \bar{G}_2 - 2\sqrt{\bar{G}_1 \bar{G}_2}\cos(\bar{\Phi}_1 - \bar{\Phi}_2) + \left(1 - \frac{\bar{G}_2}{\sqrt{\bar{G}_1 \bar{G}_2}}\cos(\bar{\Phi}_1 - \bar{\Phi}_2)\right)\delta g_1 \\ + \left(1 - \frac{\bar{G}_1}{\sqrt{\bar{G}_1 \bar{G}_2}}\cos(\bar{\Phi}_1 - \bar{\Phi}_2)\right)\delta g_2 + 2\sqrt{\bar{G}_1 \bar{G}_2}\,(\delta\varphi_1 - \delta\varphi_2)\sin(\bar{\Phi}_1 - \bar{\Phi}_2)\Bigg) \end{aligned} \tag{9}$$

where $n_{1,\omega_{cki-dat},s}$ is given by expression (32) in the Appendix and $C_{OP_a}$ is a constant at a given operating point defined by:

$$C_{OP_a} = \frac{\partial G_{1,a}}{\partial N} - \frac{1}{\bar{G}_{1,a}} \frac{\partial G_{1,a}}{\partial N} \sqrt{\bar{G}_{1,a}\bar{G}_{2,a}} \cos\left(\bar{\Phi}_{1,a} - \bar{\Phi}_{2,a}\right) + \frac{\partial \Phi_{1,a}}{\partial N} 2\sqrt{\bar{G}_{1,a}\bar{G}_{2,a}} \sin\left(\bar{\Phi}_{1,a} - \bar{\Phi}_{2,a}\right) \quad (17)$$

Equations (15) and (16) can be further simplified given that both SOAs are not operating in the high saturation region. In this case, the first term in the brackets is the most significant one, similar to [9][21], since $n_{1,\omega_{cki-dat},a}$ is a second-order perturbation term of the carrier density variation [22].

$$p_{J,cki-dat,s} = \frac{C_{OP_s}}{16} n_{1,\omega_{cki},s} p^*_{dat,s} \quad (18)$$

$$p_{J,cki-dat,m} = \frac{C_{OP_m}}{16} n^*_{1,\omega_{dat},m} p_{cki,m} \quad (19)$$

By replacing in (18) and (19) $n_{1,\omega_{cki},s}$ and $n^*_{1,\omega_{dat},m}$ by their expressions (37) and (38) given in the Appendix, we get:

$$p_{J,cki-dat,s} = -K_s \frac{p^*_{dat,s} p_{cki,s} \bar{G}_{1,s} \tau_d}{1 + j\omega_{cki}\tau_d} \quad (20)$$

$$p_{J,cki-dat,m} = -K_m \frac{p^*_{dat,m} p_{cki,m} \bar{G}_{1,m} \tau_d}{1 - j\omega_{dat}\tau_d} \quad (21)$$

where constant $K_a$ depends both on the SOAs structural parameters and the operating point of the SOA-MZI according to (40) in the Appendix and $\tau_d$ is the SOAs differential carrier lifetime.

## IV. Conversion Gain Results

### A. *Experimental conditions*

This section describes the experimental conditions of this work and presents the static experimental results.

The setting of the SOA-MZI operating point, which is defined by the bias current $I_{SOAk}$ of SOAs, the center wavelength $\lambda_{ck}$ of the sampling signal, the wavelength $\lambda_{dat}$ of the signal to be sampled and the average input powers $P_{ctrl}$ and $P_{in}$, is very critical for both architectures.

Firstly, the current $I_{SOAk}$ is chosen to be the same for both SOAs, i.e. 360 mA. This value is somewhat lower than that used previously [15], since it helps to better control the temperature and consequently obtain a more stable response of the SOA-MZI. Furthermore, the wavelength of the signals at the SOA-MZI inputs is selected to be the same for both architectures, so as to ensure a fair comparison between them. Therefore, the signal injected at SOA-MZI input port A and the one injected at input port C are chosen and kept fixed: $\lambda_{ctrl}$ at 1550 nm and $\lambda_{in}$ at 1557.4 nm, respectively. Finally, the mean power of the signal at port C is -15 dBm.

An Optical Pulse Clock (OPC) source, which is an active mode-locked laser (Pritel model UOC-E-05-20), is driven by an RF generator at a frequency $f_{ck}$ equal to 10 GHz and provides an optical pulse train of 2 ps full-width at half-maximum pulses.

The center frequency of the Optical Band-Pass Filter (OBPF) placed at each SOA-MZI output is tuned at 1557.4 nm to select the sampled signal produced by the interferometer, while removing the optical signal centered at 1550 nm. The OBPF bandwidth is chosen to be 0.7 nm in order to filter the Amplified Spontaneous Emission (ASE) noise while permitting the harmonics needed for the 39 GHz up-conversion to pass.

The phase-shifters $\Phi_{01}$ and $\Phi_{02}$ in the two arms of the SOA-MZI and the polarization controller (PC2) placed in the path of port C are adjusted so that the optical power at output port J is minimum for both architectures when there is no power at input port A.

Secondly, in order to find the $P_{ctrl}$ power that provides the highest possible linearity, a quasi-static characterization with pulses is performed for both architectures. The procedure is similar to the pump- probe static characterization in [13] with the difference that we inject an OPC signal at the appropriate port of SOA-MZI instead of a CW signal. This type of characterization incorporates the dynamics and the behavioral changes caused to the SOA-MZI response by the utilization of pulses compared to the static case.

For the Switching architecture, the electrical power recorded in an Electrical Spectrum Analyzer (ESA) is measured at 10 GHz while changing the power of the optical carrier, $P_{ctrl}$, at $\lambda_{ck,s} = \lambda_{ctrl}$ (1550 nm) at the input port A. A CW laser source provides a constant signal at the optical carrier $\lambda_{dat,s} = \lambda_{in}$ (1557.4 nm), whose power is set by an optical attenuator at $P_{in}$ = -15 dBm at SOA-MZI input port C.

Afterwards, the second derivative (SD) of the optical modulation power (OMP) of the signal at 10 GHz at SOA-MZI output ports I and J is calculated from a 5$^{th}$ order polynomial data fitting in function of the average optical power, $P_{ctrl}$, as shown in Fig. 3. When the second derivative is zero, the first derivative is in turn constant and so the SOA-MZI response is linear. The linearity of the operating point is important as it is associated to lower distortion of the sampled signal and therefore of the up-converted signal. The $P_{ctrl}$ power used for conducting the measurements that follow is chosen close to the curves' intersection point in Fig. 3 and amounts to 0.114 mW (~ - 9.5 dBm).

For the Modulation architecture, the power of the CW optical carrier at $\lambda_{dat,m} = \lambda_{ctrl}$ (1550 nm) launched at SOA-MZI input port A is controlled while being measured in a power meter. The OPC source generates a signal at $\lambda_{ck,m} = \lambda_{in}$ (1557.4 nm), whose average power is adjusted by an attenuator at $P_{in}$ = -15 dBm at SOA-MZI input port C.

According to Fig. 4, the point at which the SD of the SOA-MZI output power at port J crosses zero occurs at 0.04 mW (-14 dBm), which hence is chosen as the $P_{ctrl}$ power for the measurements taken below with the Modulation architecture.

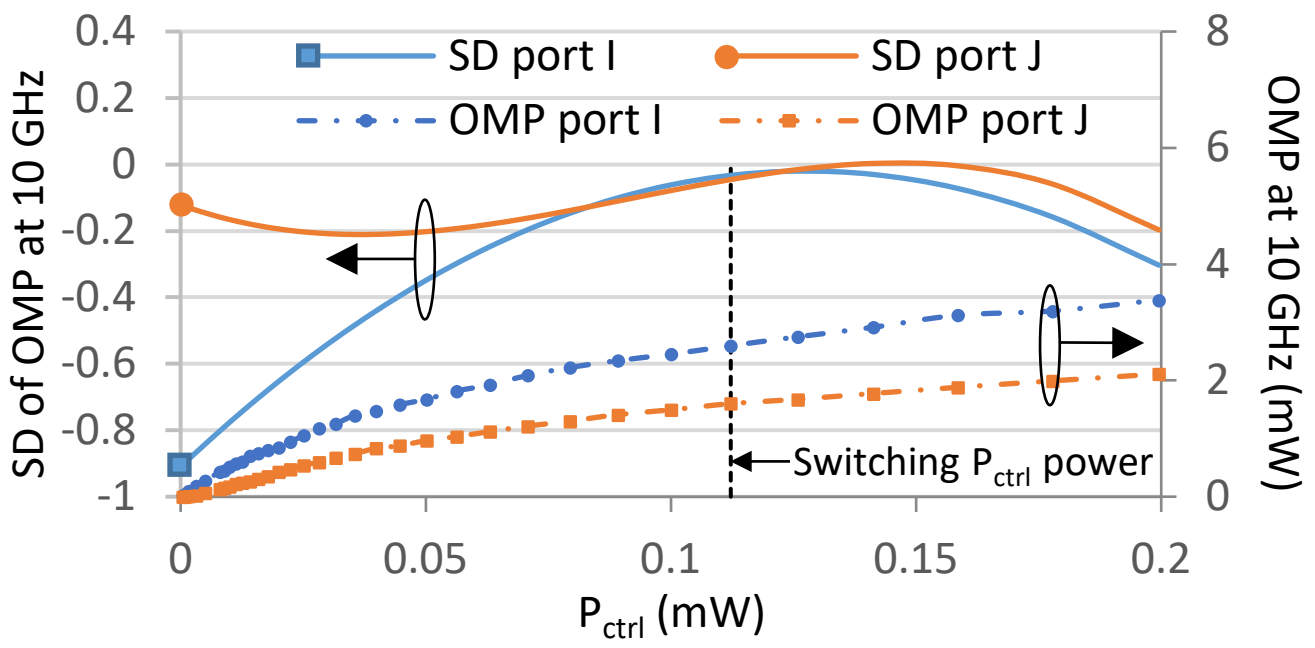

Fig. 3. Optical Modulation Power (OMP) (right) and its Second Derivative (SD) (left) at 10 GHz at SOA-MZI output ports I and J as a function of the optical power at input port A.

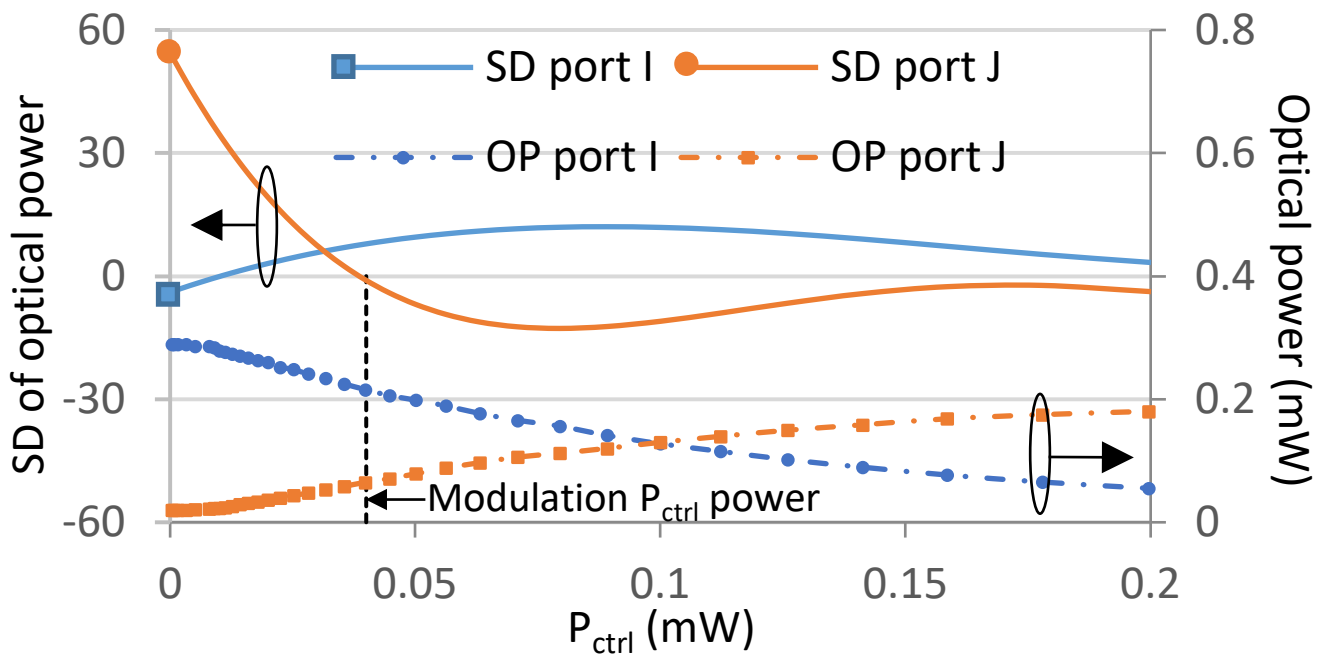

Fig. 4. Optical power (right) and its Second Derivative (SD) at SOA-MZI output ports I and J as a function of the optical power at input port A.

## B. Conversion Gain

In order to quantify the performances of the photonic microwave mixer, we have evaluated the electrical conversion gain, $CG_{cki-dat,a}$. The use of electrical conversion gain instead of optical one permits us to easily compare the small-signal analysis results with the experimental ones. The conversion gain is defined as the squared modulus of the ratio between $p_{J,cki-dat,a}$, i.e. the optical output modulation power at frequency $f_{cki-dat}$ of the optical carrier at wavelength $\lambda_{in}$, and $p^*_{dat,a}$, i.e. the optical input modulation power at frequency $f_{dat}$.

$$CG_{cki-dat,a} = \left|\frac{p_{J,cki-dat,a}}{p^*_{dat,a}}\right|^2 \tag{22}$$

Applying (20) and (21) to (22), the electrical conversion gain for the Switching and Modulation architectures is given by (23) and (24), respectively:

$$CG_{cki-dat,s} = \left|-K_s \frac{p_{cki,s}\bar{G}_{1,s}\tau_d}{1 + j\omega_{cki}\tau_d}\right|^2 \tag{23}$$

$$CG_{cki-dat,m} = \left|-K_m \frac{p_{cki,m}\bar{G}_{1,m}\tau_d}{1 - j\omega_{dat}\tau_d}\right|^2 \tag{24}$$

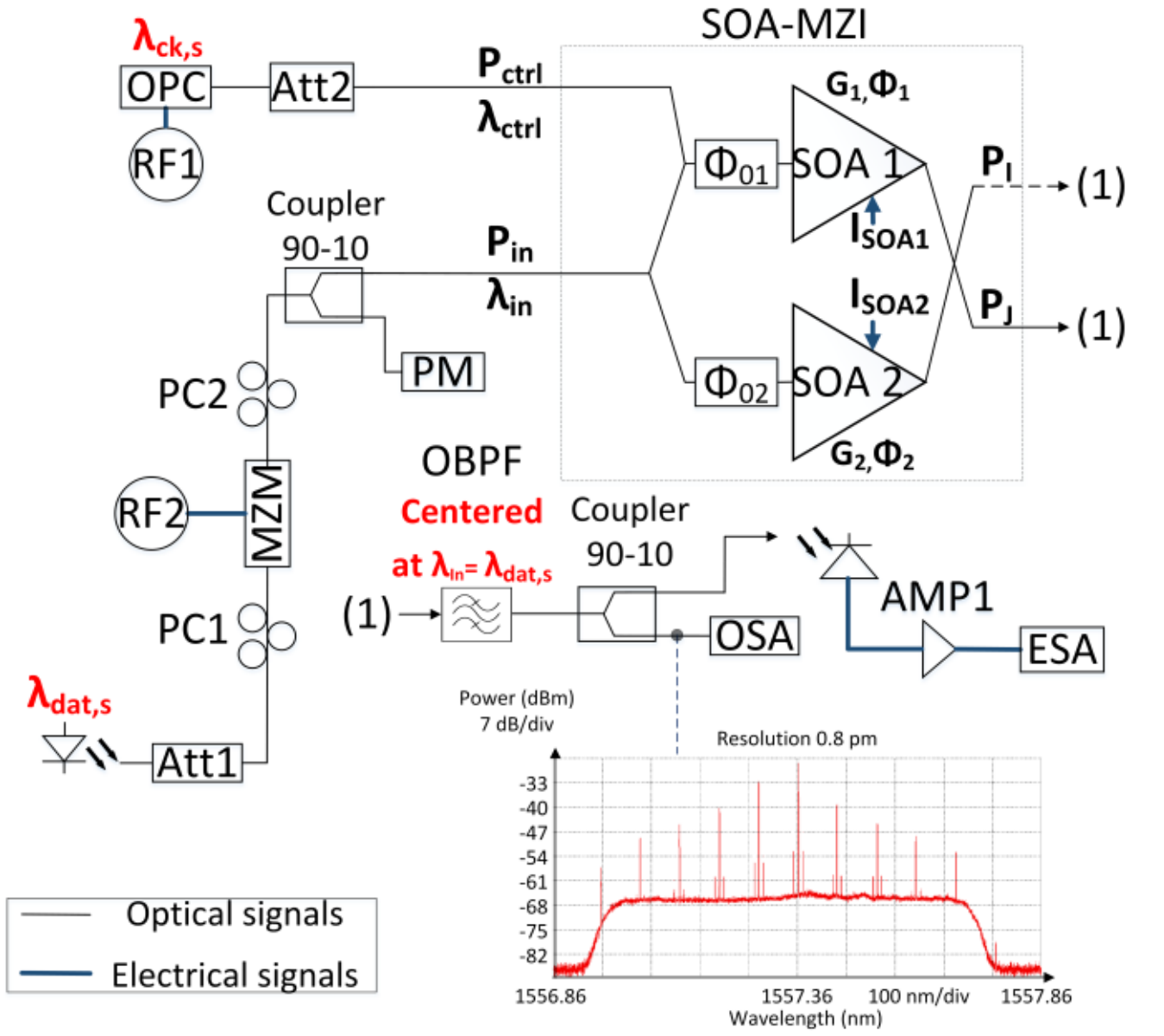

Fig. 5. *CG* measurement setup for Switching architecture. The inset depicts the output optical spectrum at the 10 % output port of the 10-90 coupler. Att: Attenuator. MZM: Mach-Zehnder Modulator. PC: Polarization Controller. OBPF: Optical Band-Pass Filter (bandwidth 0.7 nm). AMP1: RF Amplifier. OSA: Optical Spectrum Analyzer. RF1, RF2: Radio Frequency generators. ESA: Electrical Spectrum Analyzer, OPC: Optical Pulse Clock source. (1) denotes the direction of transmission from SOA-MZI output ports I and J alternately to the ESA.

Eq. (23) and (24) indicate that the *CG*s exhibit a low-pass filter response, whose cut-off frequency occurs at $f_c = \frac{1}{2\pi\tau_d}$ [17][21]. In case where $\omega_{ck1}$ is much higher than $\omega_c = 2\pi f_c$, while $\omega_{dat}$ is lower than $\omega_c$, we can make the projection that for the Switching architecture, as the order 'i' of $\omega_{cki}$ increases, the *CG* will decrease rapidly following the low-pass filter response of the carrier density. On the other hand, for the Modulation architecture, as 'i' increases, the *CG* is not subject to this drop. The derived small-signal equations show the advantage of the Modulation architecture for frequency up-conversion around higher harmonics of the sampling signal compared to the Switching one.

The *CG* has been calculated using the above small-signal equations and subsequently measured for the frequency up-conversions from $f_{dat}$ = 1 GHz to $f_{ck} - f_{dat}$ = 9 GHz ($CG_{9GHz}$) and to $4f_{ck} - f_{dat}$ = 39 GHz ($CG_{39GHz}$).

The experimental setup used to measure the *CG* is shown in Fig. 5 and Fig. 6 for the Switching and Modulation architectures, respectively. In both figures, the output spectrum of the sampled signal is given (inset) at the 10% output port of the 10-90 coupler.

The signal to be up-converted is produced for both architectures by a CW laser source that is intensity modulated by a MZM, which is driven by an RF generator RF2 at $f_{dat}$ = 1 GHz. This optically carried RF signal is then frequency converted around $f_{cki}$ frequencies. The optical signal emerging from the SOA-MZI I and J outputs is optically filtered by the previously mentioned OBPF. The filtered optical signal is monitored in an Optical Spectrum Analyzer (OSA) using the 10% port of a 90/10 coupler.

From the 90% port of the coupler, the signal is subsequently photodetected by a 75 GHz PIN photodiode whose responsivity is 0.71 A/W. The combined optical loss of the OBPF and the coupler (90% output) is 7.5 dB. The electrical output signal is then amplified by a 33 dB low noise amplifier (AMP 1).

The *CG* is defined as the difference between the electrical power in dBm referenced at the SOA-MZI output at $f_{cki} - f_{dat}$, and the electrical power in dBm referenced at the SOA-MZI input at $f_{dat}$.

In Fig. 7 we display the measured *CG* for various modulation index values of the sinusoidal signal to be sampled. Moreover, we show in dash-dot lines the theoretically predicted conversion gain derived using (23) and (24) and the involved parameters values quoted in Table I in the Appendix. For the Modulation architecture the only parameter that changes in (24) between the different up-converted signals is the power of the sampling signal harmonics, $p_{cki,m}$.

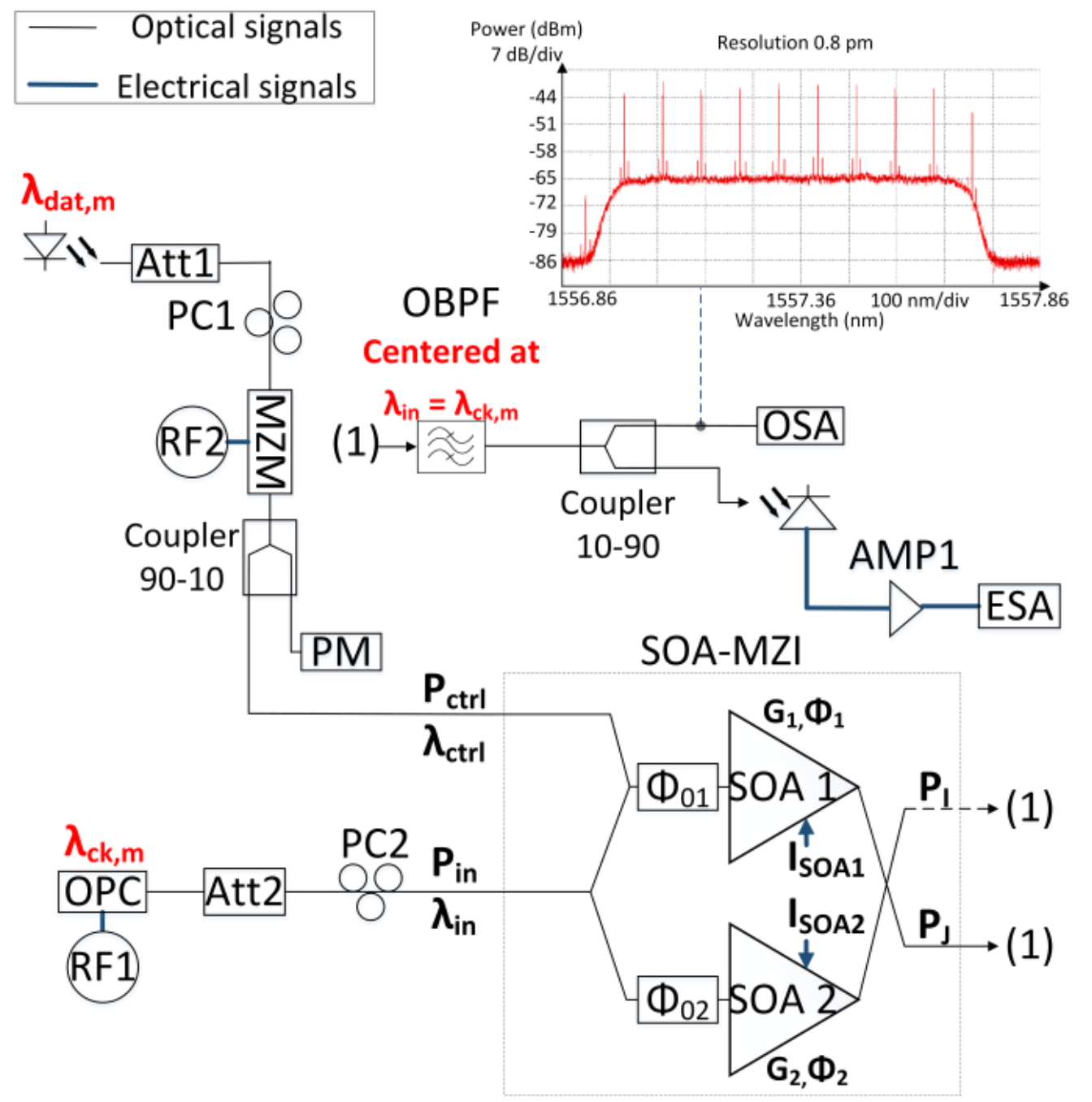

Fig. 6. *CG* measurement setup for Modulation architecture. The inset depicts the optical spectrum of the filtered sampled signal at the 10 % output port of the 10-90 coupler. The rest acronyms are defined as in Fig. 5. (1) denotes the direction of transmission from SOA-MZI output ports I and J alternately to the ESA.

There are two reasons behind the *CG* difference between the two architectures. The first reason is that the operating point is different between the Switching and Modulation architectures as $P_{ctrl}$ is -9.5 dBm for the former and -14 dBm for the latter. The second reason stems from the fact that we have interchanged the sinusoidal signal to be sampled with the mode-locked sampling signal at the input of the SOA-MZI so that the frequency components that participate in the $f_{dat} =$ 1 GHz to $f_{ck1-dat} =$ 9 GHz frequency conversion are lightly filtered by the SOA-MZI for the Switching architecture as $f_{ck}$ at 10 GHz lies outside the XPM bandwidth ($f_c = 6\,GHz$, given in the Appendix), whereas they are not filtered for the Modulation architecture as $f_{dat}$ at 1 GHz lies inside the XPM bandwidth. At $f_{ck4-dat} =$ 39 GHz the frequency components that participate in the $f_{dat} =$ 1 GHz to $f_{ck4-dat} =$ 39 GHz frequency conversion are strongly filtered by the SOA-MZI for the Switching architecture as $4f_{ck}$ at 40 GHz lies outside the XPM bandwidth, whereas they are not filtered for the Modulation architecture as $f_{dat} =$ 1 GHz lies inside the XPM bandwidth. This explains the increased robustness of the Modulation scheme against the Switching scheme with respect to *CG* as the up-conversion frequency increases.

Fig. 7 shows a good agreement between measured and theoretically calculated CG curves, which confirms the validity and accuracy of the theoretical approach and of its outcomes. Furthermore, it highlights the usability of (23) and (24) in the design, characterization and optimization of the SOA-MZI photonic mixer. The difference in *CG* between the frequency conversion to 9 GHz and 39 GHz is smaller for the Modulation architecture than for the Switching one. Therefore, although *CG* is higher for the frequency conversion to 9 GHz when employing the Switching architecture, it is more advantageous to employ the Modulation architecture for the frequency conversion to 39 GHz.

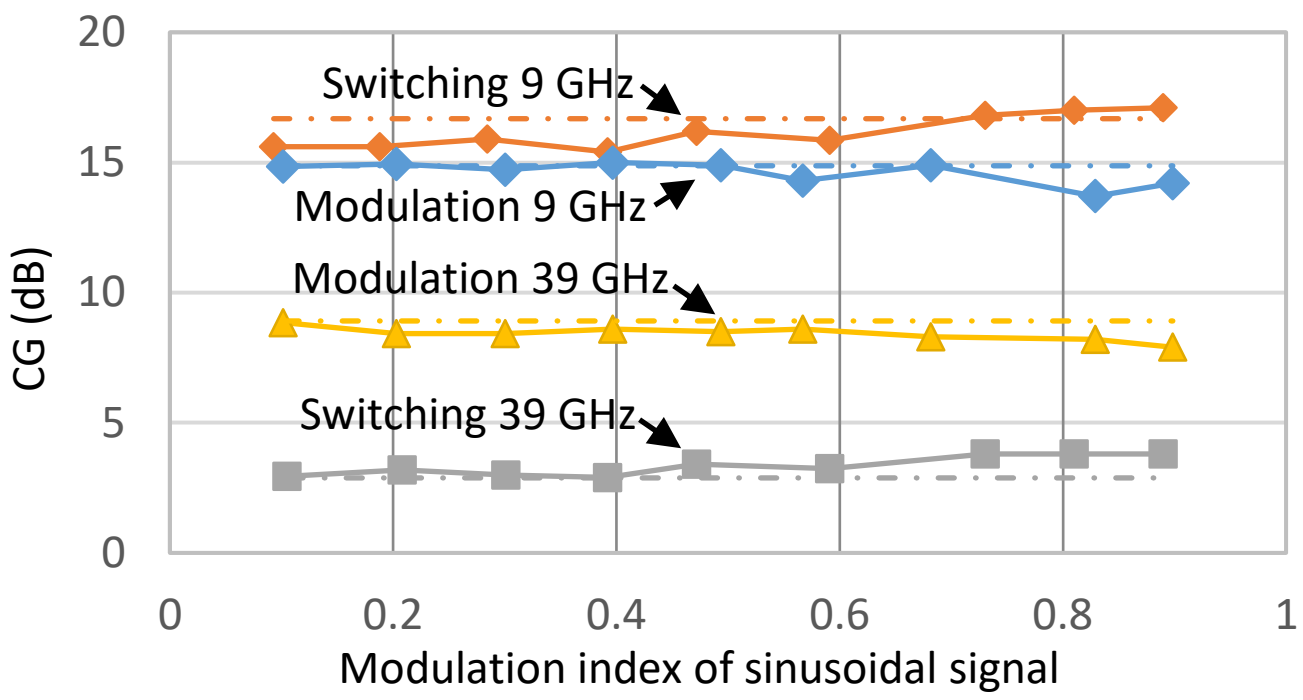

Fig. 7. Comparison between *CG* of the Switching and Modulation architectures: Measured (solid lines) and theoretically calculated (dash-dot lines) *CG* values for frequency up-conversion at 9 GHz and 39 GHz as a function of the modulation index of $P_{dat}$.

Moreover, the *CG* attained by the SOA-MZI photonic mixer does not change significantly with the modulation index variation. Eq. (23) and (24) support this fact as *CG* is not a function of $p_{dat,a}$, which in turn is proportional to the modulation index, i.e. $p_{dat,a} = m_{dat,a}\bar{P}_{dat,a}$, where $m_{dat,a}$ is the modulation index of the sinusoidal signal.

Furthermore with respect to the SOA-MZI parameters influence on *CG*, from Eq. (23), Eq.(24) and Eq. (40) we observe that the electrical *CG* is inversely proportional to the square of the saturation power, $P_{sat}$, and to the carrier lifetime $\tau$, both of which depend on the SOA physical properties [9]. Additionally, we remark, considering the other parameters constant (including the ratio $\frac{\bar{G}_{1,a}}{\bar{G}_{2,a}}$), that $\bar{G}_{1,a}$is proportional to the 4th power of CG (in the electrical domain). Nevertheless, for a given SOA-MZI, $\bar{G}_{1,a}, \tau_d$, $K_a$, $p_{cki,a}$, $P_{sat}$and $\tau$ are interdependent and associated to its operating condition. Therefore the effect of each critical SOA-MZI parameter depends in any case on the exact operating point of the SOA-MZI.

It should be emphasized that in the case where either the input sampling signal or the output sampled signal has to be transmitted over a long distance through an optical fiber, the problem of fiber chromatic distortion must be addressed. One possible solution to compensate for the chromatic dispersion and overcome the associated problem is through using a Dispersion Compensating Fiber (DCF) of appropriate length just before the electro-optic conversion stage.

In addition, it is worth noting that this system could be applicable to CWDM and DWDM systems provided that $if_{ck} \pm f_{dat}$ is smaller than the channel spacing of the corresponding grid.

## V. Up-conversion of complex-modulated data Experimental results

The experimental setup of the SOA-MZI photonic frequency up-converter using complex modulation formats is shown in Fig. 8 for the Switching architecture and in Fig. 9 for the Modulation architecture. QPSK and 16-QAM signals are generated by an Arbitrary Waveform Generator (AWG) at a carrier frequency $f_c$ = 0.750 GHz. The electrical signal from the AWG modulates an optical carrier, via a MZM, which is injected at the SOA-MZI.

In the same way as in the *CG* experimental setups, the sampled output signal at ports I and J is filtered by the OBPF, photodetected and amplified by the low noise amplifier (AMP 1).

Due to the Digital Sampling Oscilloscope (DSO) limited bandwidth (about 1.5 GHz), the up-converted RF signals at 9.250 GHz and 39.250 GHz at the SOA-MZI output are down-converted to IF signals at 450 MHz by an electrical mixer. The electrical power of the local oscillator signal at $f_{RF2}$ is 13 dBm. The electrical mixer is characterized by a 10 dB conversion loss. Before the DSO, the up-converted signal is filtered by a 1 GHz electrical filter and amplified by a 40 dB electrical amplifier (AMP 2).

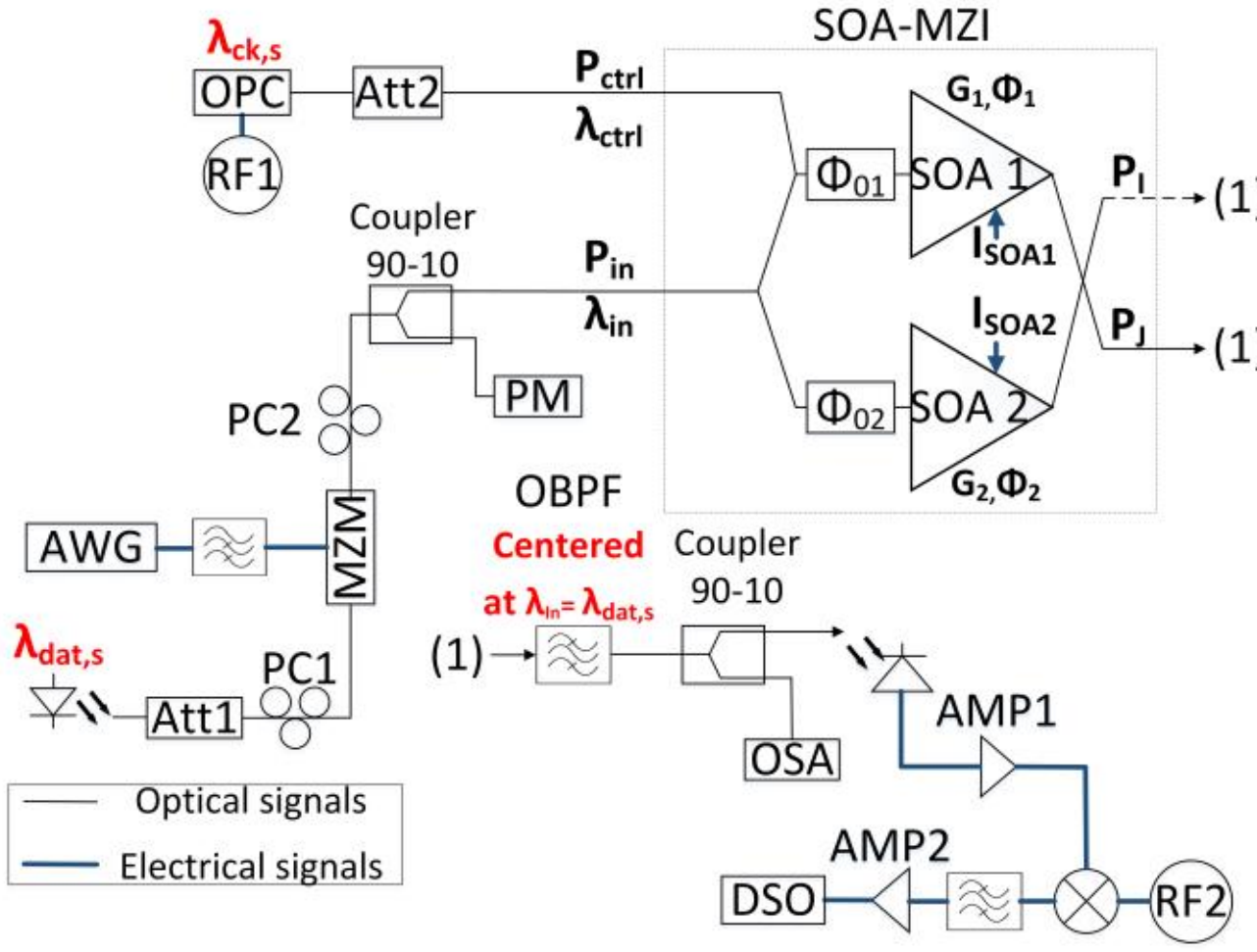

Fig. 8. Complex modulation measurement setup for Switching architecture. AWG: Arbitrary Waveform Generator. AMP1, AMP2: RF Amplifiers. DSO: Digital Sampling Oscilloscope. The rest acronyms are defined as in Fig. 5. (1) denotes the direction of transmission from SOA-MZI output ports I and J alternatively to the DSO.

The quality of the up-converted signals is assessed by measuring the Error Vector Magnitude (EVM) [24] via a Vector Signal Analyzer software. This metric is related to the Bit Error Rate (BER), and the acceptable limit is defined as the value that provides an equivalent BER of $3.8 \cdot 10^{-3}$, which guarantees error-free performance after applying Forward Error Correction (FEC) techniques [25]. This limit is different for QPSK and 16-QAM modulation formats as indicated by a dashed line in Figs. 10 and 11. These figures show that the up-converted signals at 9.25 GHz and 39.25 GHz using the Modulation architecture have very similar quality in terms of EVM, in contrast to significant differences observed when using the Switching architecture. This is explained by the small-signal equation analysis presented in Section III and stems from the different principle of operation of the two architectures. Using the Modulation architecture, we can perform up-conversion even at high harmonics without significant degradation of the up-converted signal quality.

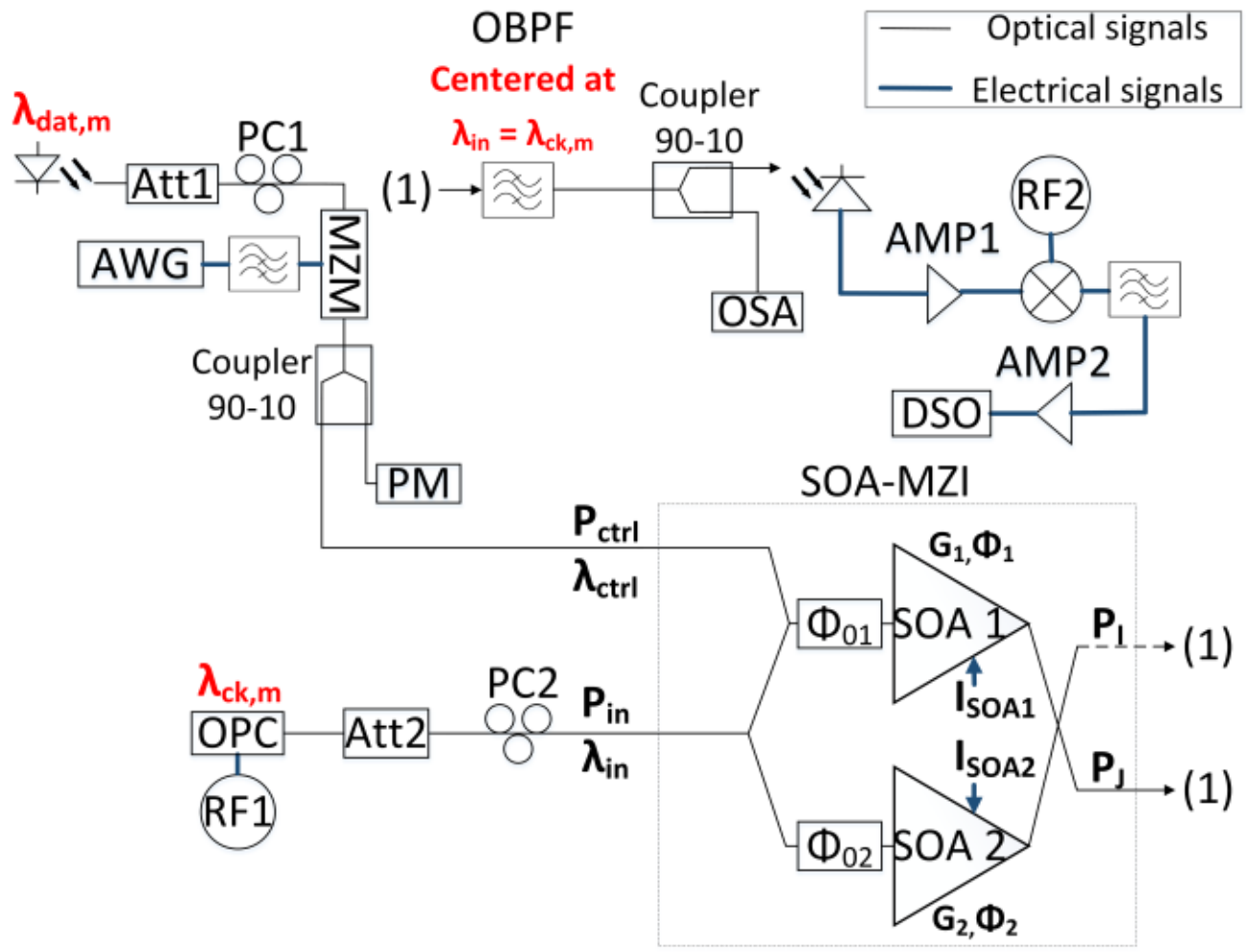

Fig. 9. Complex modulation measurement setup for Modulation architecture. PM: Power Meter. The rest acronyms are defined as in Fig. 5. (1) denotes the direction of transmission from SOA-MZI output ports I and J alternatively to the DSO.

In Fig.10, for QPSK-modulated data, a baud rate of 512 MBaud is achieved for frequency conversions from 0.75 GHz to 9.25 GHz and from 0.75 GHz to 39.25 GHz under the FEC limit of $3.8 \cdot 10^{-3}$ when employing the Modulation architecture. On the other hand, the Switching architecture performance deteriorates rapidly as the baud rate increases, reaching the limit of EVM ~ 35% for the conversion towards 39.25 GHz and approximating EVM = 25% for the conversion towards 9.25 GHz at a baud rate of 256 MBaud. Measurements at a baud rate of 512 MBaud for the Switching architecture for the conversion to 39.25 GHz and even to 9.25 GHz have resulted in significantly distorted constellation diagrams and EVM values higher than 37%, and therefore they have not been included in Fig.10. Comparing the inset constellation diagrams in Fig.10 it can be deduced that the QPSK signal for the Switching architecture between 9.25 GHz and 39.25 GHz is significantly deteriorated even for a lower baud rate, whereas comparing the inset constellation diagrams for the Modulation architecture it can be inferred that the QPSK signal for the Modulation architecture between 9.25 GHz and 39.25 GHz remains almost unchanged and clear.

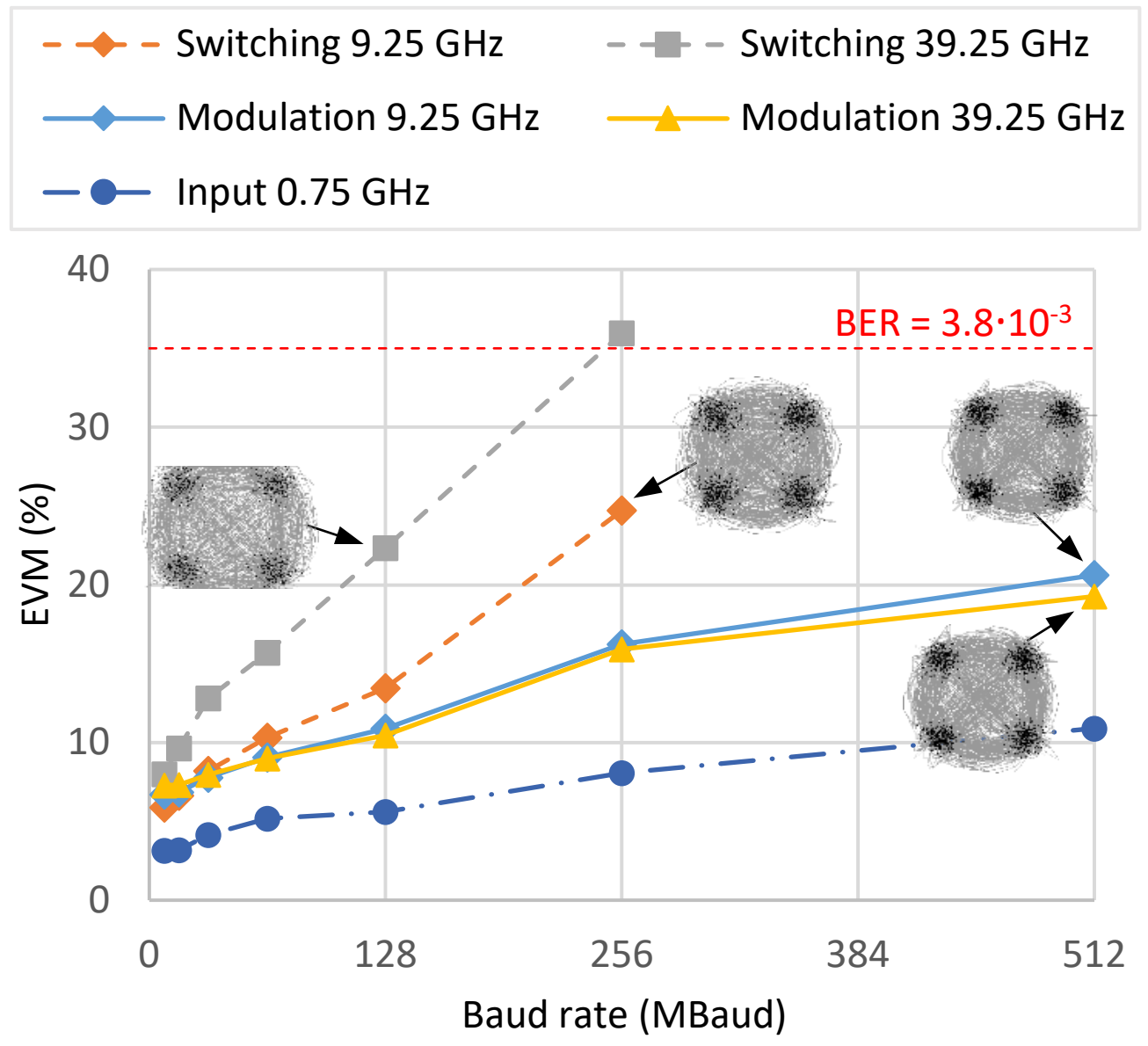

Fig. 10. EVM vs. baud rate comparison between Switching and Modulation architectures for frequency up-conversion of QPSK-modulated data at 9.25 GHz and 39.25 GHz. The EVM of the input signal at 0.75 GHz to be up-converted is plotted as a reference. The insets depict constellation diagrams at specific points. EVM acceptable limit is indicated by the horizontal dashed line.

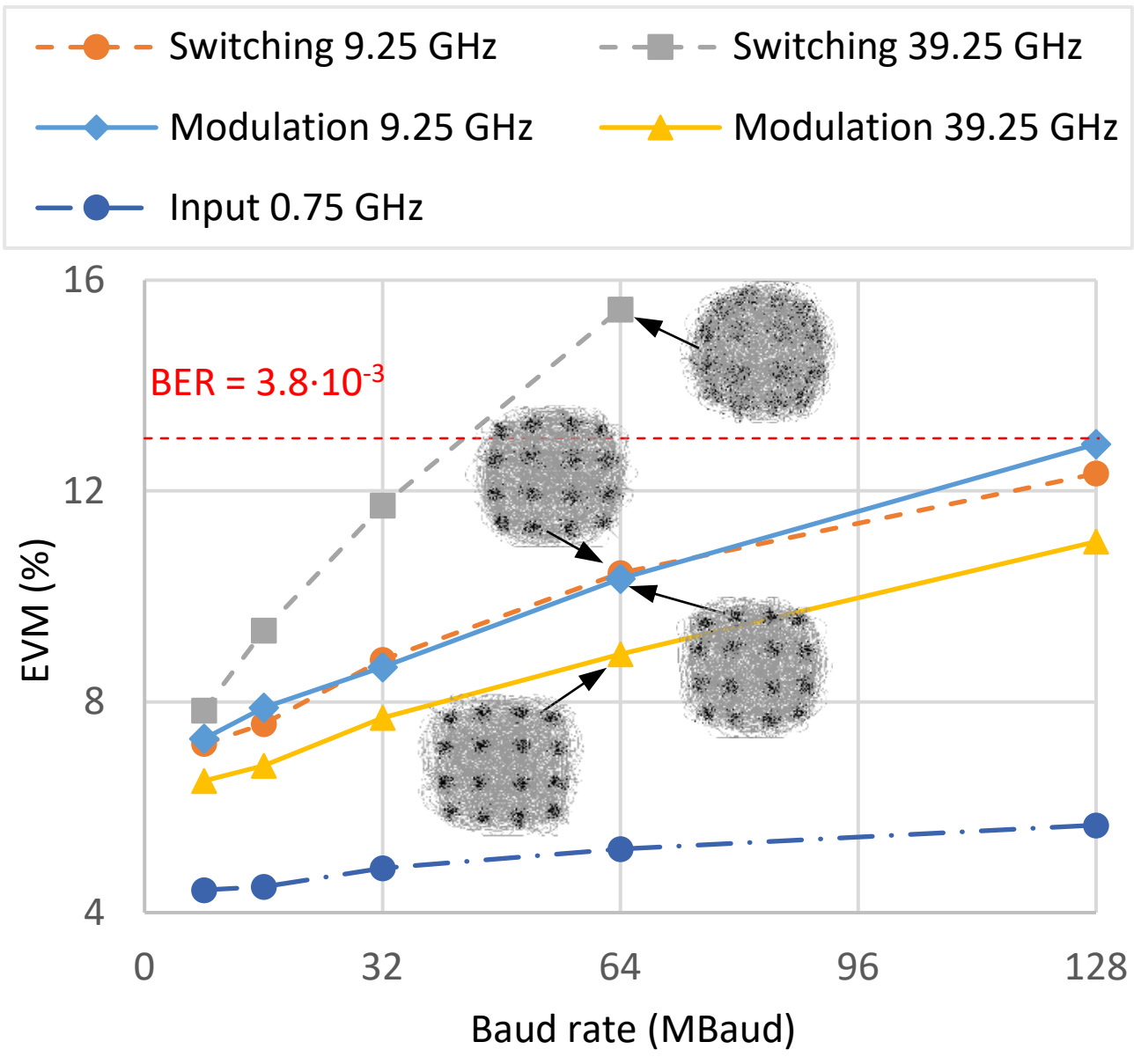

Fig. 11. EVM vs. baud rate comparison between Switching and Modulation architectures for frequency up-conversion of 16-QAM modulated data at 9.25 GHz and 39.25 GHz. The EVM of the input signal at 0.75 GHz to be up-converted is plotted as a reference The insets depict constellation diagrams at specific points. EVM acceptable limit is indicated by the horizontal dashed line.

In Fig. 11, a baud rate of 128 Mbaud is achieved for 16-QAM modulated data with the Modulation architecture for frequency conversions from 0.75 GHz to 9.25 GHz and from 0.75 GHz to 39.25 GHz under the FEC limit of $3.8 \cdot 10^{-3}$. The same holds for the Switching architecture for frequency conversions from 0.75 GHz to 9.25 GHz. Conversely, for frequency conversions from 0.75 GHz to 39.25 GHz only a baud rate up to 32 Mbaud is acceptable. It is worth noting that if we observe more closely the 16-QAM constellation diagrams in Fig. 11, we realize that their shape is not perfectly squared and that the demodulated symbols corresponding to the four diagonal outer points are misplaced towards the center, thus giving the impression that the constellation diagrams are compressed to the center of the constellation. This is rather a sign of amplitude distortion than of phase distortion.

16-QAM modulation is more demanding as the Euclidean distance between symbols in the constellation diagram is smaller [26]. Thus, the highest acceptable baud rate for QPSK is 512 MBaud (bit rate equal to 1 Gb/s), while it is 128 MBaud (bit rate equal to 512 Mb/s) for 16-QAM.

## VI. Conclusion

We have presented a theoretical and experimental analysis of the performance differences between Switching and Modulation architectures of a SOA-MZI photonic mixer used for frequency up-conversion purposes. In the theoretical part, a closed-form expression has been derived for the conversion gain, which provides qualitative insight into this metric of the two architectures as higher harmonics of the sampling signal participate in the frequency conversion. In the experimental part, a conversion gain equal to 16 dB is achieved for the 1 GHz to 9 GHz conversion when employing the Switching architecture, while a conversion gain equal to 9 dB is achieved for the 1 GHz to 39 GHz conversion when employing the Modulation architecture. The matching between the experimental and theoretical results for the conversion gain is very good. The conversion of complex-modulated data signals has also been realized by the photonic microwave mixer. We have thus obtained a sufficiently low EVM for QPSK and 16- QAM modulations when employing the Modulation architecture for both 0.75 GHz to 9.25 GHz and 0.75 GHz to 39.25 GHz frequency conversion. This allows to support a data bit rate equal to 1 Gbps and 512 Mbps for QPSK and 16-QAM, respectively.

## Appendix

### A. *Carrier Density Variation*

The gain of SOAk as well as the partial derivative of the gain with respect to carrier density are given by the following equations [9]:

$$G_k(N_k) = e^{\Gamma a_k (N_k - N_0)L + a_{int}L} \tag{25}$$

$$\frac{\partial G_k}{\partial N_k} = \Gamma a_k L \bar{G}_k \tag{26}$$

with $L$ the active region length, $\Gamma$ the optical confinement factor in the active area, $a_k$ the peak-gain coefficient, $N_0$ the carrier density at transparency, and $a_{int}$ the internal losses, of the SOAs.

The phase shift of the optical signal induced by SOAk as well as its partial derivative with respect to the carrier density are given by [23]:

$$\Phi_k = -\frac{\alpha_H}{2} ln(G_k) = -\frac{\alpha_H}{2}[\Gamma a_k(N_k - N_0)L + a_{int}L] \quad (27)$$

$$\frac{\partial \Phi_k}{\partial N_k} = -\frac{\alpha_H}{2}\Gamma a_k L \quad (28)$$

where $\alpha_H$ is the linewidth enhancement factor (Henry's factor).

Taking into account (27) and assuming identical $\alpha_H$ and $a_{int}$ values for both SOAs, the differential phase shift between the two arms of the SOA-MZI equals:

$$\bar{\Phi}_{1,a} - \bar{\Phi}_{2,a} = -\frac{\alpha_H}{2} ln\left(\frac{\bar{G}_{1,a}}{\bar{G}_{2,a}}\right) \quad (29)$$

Assuming for simplicity that the differential carrier lifetime $\tau_d$ is approximately the same for SOA1 between Switching and Modulation architectures, the first-order variation imposed on the carrier density is given by [9]:

$$n_{1,\omega_{cki},a} = -\frac{p_{cki,a}\frac{\partial R_{1,cki,a}}{\partial P_{1,cki,a}}\tau_d}{\eta_a(1 + j\omega_{cki}\tau_d)} \quad (30)$$

$$n^*_{1,\omega_{dat},a} = -\frac{p^*_{dat,a}\frac{\partial R_{1,dat,a}}{\partial P_{1,dat,a}}\tau_d}{\kappa_a(1 - j\omega_{dat}\tau_d)} \quad (31)$$

where $\eta_a$ and $\kappa_a$ are the attenuation coefficients of $p_{cki,a}$ and $p^*_{dat,a}$ because of the intermediate couplers between the SOA-MZI input ports and SOA1 input according to Fig.2.

The second-order density variation of the carrier density is given by [9]:

$$n_{1,\omega_{cki-dat},a} = -\left(n_{1,\omega_{cki},a}\frac{p^*_{dat,a}}{\kappa_a}\frac{\partial^2 R_{1,dat,a}}{\partial N_1 \partial P_{1,dat,a}} + n^*_{1,\omega_{dat},a}\frac{p_{cki,a}}{\eta_a}\frac{\partial^2 R_{1,cki,a}}{\partial N_1 \partial P_{1,cki,a}}\right) \times \frac{\tau_d}{2(1 + j(\omega_{cki} - \omega_{dat})\tau_d)} \quad (32)$$

$R_{1,cki,a}$ and $R_{1,dat,a}$ are the recombination rates due to the amplification of the optical carriers at angular frequencies $\omega_{cki}$ and $\omega_{dat}$, respectively, and are given by the following equations [9]:

$$R_{1,cki,a} = \frac{P_{1,cki,a}\lambda_{cki,a}}{hc}\frac{\Gamma g_{m,cki,a}}{g_{n,cki,a}}\frac{\bar{G}_{1,a} - 1}{wdL} \quad (33)$$

$$R_{1,dat,a} = \frac{P_{1,dat,a}\lambda_{dat,a}}{hc}\frac{\Gamma g_{m,dat,a}}{g_{n,dat,a}}\frac{\bar{G}_{1,a} - 1}{wdL} \quad (34)$$

where $h$ is Plank's constant, $c$ is the speed of light in vacuum, $g_{m,cki,a}$ and $g_{n,cki,a}$ are the SOAs material gain and net gain at wavelength $\lambda_{cki,a}$, $g_{m,dat,a}$ and $g_{n,dat,a}$ are the SOAs material gain and net gain at wavelength $\lambda_{dat,a}$, and $w$ and $d$ are the SOAs width and height of the active area, respectively.

Using the approximations $\Gamma g_{m,cki,a} \approx g_{n,cki,a}$, $\Gamma g_{m,dat,a} \approx g_{n,dat,a}$, $G_{1,a} - 1 \approx G_{1,a}$ and $\lambda \approx \lambda_{ck,s} \approx \lambda_{dat,s}$ as in [9], we obtain:

$$\frac{\partial R_{1,cki,a}}{\partial P_{1,cki,a}} = \frac{\partial R_{1,dat,a}}{\partial P_{1,dat,a}} = K\bar{G}_{1,a}(N_1) \quad (35)$$

where $K$ is a constant defined by:

$$K = \frac{\lambda}{hcwdL} \quad (36)$$

Applying (35) to (30) and (31), we obtain:

$$n_{1,\omega_{cki},a} = -\frac{p_{cki,a}K\bar{G}_{1,a}(N_1)\tau_d}{\eta_a(1 + j\omega_{cki}\tau_d)} \quad (37)$$

$$n^*_{1,\omega_{dat},a} = -\frac{p^*_{dat,a}K\bar{G}_{1,a}(N_1)\tau_d}{\kappa_a(1 - j\omega_{dat}\tau_d)} \quad (38)$$

### B. *Saturation power*

The saturation power $P_{sat}$ of SOA1 is defined in [9]:

$$P_{sat} = \frac{hcwd}{\lambda \Gamma a_1 \tau} \quad (39)$$

where $\tau$ is the carrier lifetime. Finally, $K_a$ is a constant defined by:

$$K_a = \frac{KC_{OP_a}}{32} \xRightarrow{(17),(26),(28),(29),(36),(39)}$$

$$K_a = \frac{\bar{G}_{1,a}}{32P_{sat}\tau}\left[1 - \sqrt{\frac{\bar{G}_{2,a}}{\bar{G}_{1,a}}}\left[\cos\left(-\frac{\alpha_H}{2} ln\left(\frac{\bar{G}_{1,a}}{\bar{G}_{2,a}}\right)\right) + \alpha_H \sin\left(-\frac{\alpha_H}{2} ln\left(\frac{\bar{G}_{1,a}}{\bar{G}_{2,a}}\right)\right)\right]\right] \quad (40)$$

### C. *SOA-MZI model parameters*

The differential carrier lifetime $\tau_d$ is estimated by measuring the SOA-MZI bandwidth using a pump-probe technique [13] and taking into account that the measured cut-off frequency of 6 GHz equals $\frac{1}{2\pi\tau_d}$ [17][21].

The optical power, $p_{cki,a}$, required for the small-signal analysis calculations is derived from RF spectra for the Switching and Modulation architectures. These spectra were taken after photodetection and amplification of the OPC signal at the average power and center wavelength specified in Section IV. We know that the photodetected current is defined as:

$$I_{ph} = p_{cki,a}r \quad (41)$$

where $r$ (0.71 A/W) is the responsitivity of the PIN photodiode. The electrical power for harmonic '$i$' measured at the ESA is equal to:

$$p_{eli,a} = G_{amp} I_{rms}^2 Z = G_{amp} \frac{I_{ph}^2}{2} Z \quad (42)$$

where $G_{amp}$ (33 dB) is the gain of the RF amplifier, $I_{rms}$ is the root mean square value of the photodetected currect and Z (50 Ω) is the load resistance of the photodetector. Therefore, applying (41) to (42) and solving for the optical power $p_{cki,a}$ we obtain:

$$p_{cki,a} = \frac{1}{r}\sqrt{\frac{2p_{eli,a}}{G_{amp}Z}} \quad (43)$$

Table I lists the parameter values of SOA-MZI small-signal model used for the *CG* calculation.

TABLE I
PARAMETER VALUE FOR SOA-MZI SMALL-SIGNAL MODEL

| Symbol | Parameter | Value | Unit |
|---|---|---|---|
| $a_H$ | Linewidth enhancement factor | 4 | |
| $\tau_d$ | Differential carrier lifetime | 26.5 | ps |
| $\tau$ | Carrier lifetime | 70 | ps |
| $P_{sat}$ | SOA1 saturation power (Both architectures) | 15 | dBm |
| $\bar{G}_{1,s}$ | SOA1 Gain (Switching) | 24.8 | dB |
| $\bar{G}_{1,m}$ | SOA1 Gain (Modulation) | 26.1 | dB |
| $\bar{G}_{2,a}$ | SOA2 Gain (Both architectures) | 28 | dB |
| $p_{el1,s}$ | Electrical power at 10 GHz (Switching) | -8 | dBm |
| $p_{el4,s}$ | Electrical power at 40 GHz (Switching) | -11 | dBm |
| $p_{el1,m}$ | Electrical power at 10 GHz (Modulation) | -18 | dBm |
| $p_{el4,m}$ | Electrical power at 40 GHz (Modulation) | -24 | dBm |
| $p_{ck1,s}$ | OPC OMP power at 10 GHz (Switching) | -11 | dBm |
| $p_{ck4,s}$ | OPC OMP power at 40 GHz (Switching) | -12.5 | dBm |
| $p_{ck1,m}$ | OPC OMP power at 10 GHz (Modulation) | -16 | dBm |
| $p_{ck4,m}$ | OPC OMP power at 40 GHz (Modulation) | -19 | dBm |
| $\eta_s$ | Attenuation coefficient of $p_{cki,s}$ | 2 | |
| $\eta_m$ | Attenuation coefficient of $p_{cki,m}$ | 4 | |
| $\kappa_s$ | Attenuation coefficient of $p_{dat,s}$ | 4 | |
| $\kappa_m$ | Attenuation coefficient of $p_{dat.m}$ | 2 | |

**Dimitrios Kastritsis** was born in Thessaloniki, Greece, in 1988. He received the Diploma in Electrical and Computer Engineering in 2015 from Democritus University of Thrace (DUTH), Xanthi, Greece. He is currently working toward the Ph.D. degree in photonics, optoelectronics, and telecoms in the École Nationale d'Ingénieurs de Brest (ENIB), France. His research interests include all-optical sampling based on a Semiconductor Optical Amplifier Mach-Zehnder Interferometer (SOA-MZI) for analog applications.

**Thierry Rampone** was born in Cavaillon, France, in 1969. He received the Engineering degree from the École Nationale d'Ingénieurs de Brest, France, in 1992, and the Diplôme d'Études Approfondies (D.E.A.) and the Ph.D. degree from the Université de Bretagne Occidentale, Brest, France, in 1992 and 1997, respectively.
His doctoral study principally concerned optical switches based on semiconductor optical amplifiers. In 1998, he joined RESO Laboratory, École Nationale d'Ingénieurs de Brest (ENIB), France, as an Associate Professor. His research interests include all-optical processing, functional elements for radio-over-fiber systems and microwave photonics mainly based on semiconductor optical amplifiers.

**Kyriakos Zoiros** was born in Thessaloniki, Greece, in 1973. He received the Diploma in Electrical and Computer Engineering in 1996 and the Ph.D. in optical communications in 2000, both from National Technical University of Athens (NTUA). He is currently an Associate Professor in the Department of Electrical and Computer Engineering, Democritus University of Thrace, Xanthi, Greece. During Spring 2009 he was on leave at the Optical Communications Research Group, University of Limerick, Ireland. During Fall 2013 he was on leave at the Laboratoire en Sciences et Techniques de l'Information, de la Communication et de la Connaissance (Lab-STICC), École Nationale d'Ingénieurs de Brest (ENIB), France. He has authored or co-authored more than 100 international journal and conference papers and 5 book chapters. His current research interests include all-optical signal processing, semiconductor optical amplifiers, microwave photonics, microring resonators and computational photonics.

**Ammar Sharaiha** was born in Amman, Jordan, in 1956. He received the M. Sc. and Ph.D degrees from the University of Rennes I, France in 1981 and 1984 respectively. He joined the École Nationale d'Ingénieurs de Brest (ENIB), France as Associate Professor (Maître de Conférences) in 1989; later, as Full Professor (Professeur des Universités) in 2001. He was head of Research at ENIB from 2010-2014, co-head of research team in Lab-STICC (UMR CNRS) from 2012 to 2016 and head of electronic research department from 2010-2016. He is author or co-author of more than 150 international publications or communications. Pr. Sharaiha's current research interests are mainly in the areas of high-speed optical-communication systems, microwave and photonics for 60GHz RoF applications using semiconductor-optical amplifiers.